# TRAVELLING WAVE SOLUTIONS OF A SIMPLE NON-LOCAL BURGERS-LIKE EQUATION

Alex Veksler[1] and Yair Zarmi[1,2]
Ben-Gurion University of the Negev, Israel
[1]Department of Physics, Beer Sheva, 84105
[2]Department of Energy & Environmental Physics
Jacob Blaustein Institute for Desert Research, Sede-Boqer Campus, 84990

## Abstract

A Lax pair consisting of the Forsyth-Hopf-Cole transformation and a linear differential (in time) - difference (in continuous space) equation generates the whole hierarchy of the Burgers equation and admits exact moving front solutions. This model mimics finite-distance spatial correlations or nearest neighbor interactions in a physical situation

Interest in semi-discrete dynamical systems has been growing in recent years. The obvious case is that of descretized versions of continuous dynamical systems (as occurs, for example in numerical simulations of continuous systems). Well-known examples are systems of coupled oscillators [1], systems of Josephson Junctions (see, E.g., [2-4]), complex networks [5] and compartmentalized granular gases [6, 7]. These systems are described by a continuous time variable and a discrete spatial variable. The fundamental aspects of the analysis of such systems have been discussed in [8-11]. In [8], the pole-expansion of solutions of PDE's was shown to correspond formally to the motion of one-dimensional particles under simple two-body potentials. In [9], the Lax pair method is applied to the class of nonlinear differential-difference equations associated with the matrix Schrödinger spectral problem. In [10], a discretization procedure is developed for integrable nonlinear evolution equations, and is applied to the discrete Burgers hierarchy. In [11], inhomogeneous Burgers lattices, which are solved by the discrete Cole-Hopf transformation, are analyzed when subjected to additive random noise. In [12], traveling wave solutions are found for bistable differential-difference equations with spatially periodic diffusion. In this note, we analyze and solve a semi-discrete dynamical system, for which the evolution equation is continuous in time, but is a difference equation in a *continuous* spatial variable. In a rough sense, this approach mimics finite-distance spatial correlations or nearest neighbor interactions in a physical situation. The equation is a simple extension of the Burgers equation (related to the one-dimensional propagation of shock waves in a fluid). It admits traveling wave solutions and serves as a generator of the whole usual Burgers hierarchy.

The system is defined by a Lax pair [13] that is comprised of the Forsyth-Hopf-Cole transformation (FHCT) [14–16]

$$y_x = u(t,x)y \tag{1}$$

and a differential-difference equation

$$y_t(t,x) = y(t,x+\lambda) - y(t,x) = (T_\lambda - 1)y(t,x) \tag{2}$$

$T_\lambda$ is the operator affecting the discrete spatial transformation:

$$T_\lambda f(t,x) = f(t,x+\lambda) \tag{3}$$

Consistency between Eqs. (1) and (2), is obtained by applying $\partial_t$ to Eq. (1) and $\partial_x$ to Eq. (2), and requiring that the results coincide, leading to a highly nonlinear integro-difference equation:

$$u_t = \frac{\partial}{\partial x}\left\{\exp\left(\int_x^{x+\lambda} u(t,\xi)d\xi\right)\right\}$$
$$= \{u(t,x+\lambda) - u(t,x)\}\exp\left(\int_x^{x+\lambda} u(t,\xi)d\xi\right) \tag{4}$$

The solution of Eq. (4) is found through its Lax pair representation. In terms of the initial data for $y(t,x)$, Eq. (2) is solved by

$$y(t,x) = e^{(T_\lambda-1)t} y_0(x) = e^{-t}\sum_{n\geq 0}\frac{t^n}{n!}y_0(x+n\lambda) \qquad (y_0(x) = y(t=0,x)) \tag{5}$$

The solution of Eq. (2) may be also written in terms of the Fourier spectrum of the initial data for $y$:

$$y(t,x) = \frac{e^{-t}}{\sqrt{2\pi}} \int_{-\infty}^{\infty} \tilde{y}_0(k) e^{ikx + \exp[i\lambda k]t} \, dk \tag{6}$$

In Eq. (6), $\tilde{y}_0(k)$ is the coefficient of the Fourier component of the initial data

$$\tilde{y}_0(k) = \frac{1}{\sqrt{2\pi}} \int_{-\infty}^{\infty} y_0(x) e^{-ikx} \, dx \tag{7}$$

The solution for $u(t,x)$ is then obtained via Eq. (1).

To see how solutions, given by Eq. (6), behave, consider the case of $\tilde{y}_0(k)$ that is real and has a narrow peak at $k=0$ (e.g., a narrow Gaussian in $k$, centered at $k=0$). We re-write Eq. (6) in the form

$$y(t,x) = \frac{1}{\sqrt{2\pi}} \int_{-\infty}^{\infty} \exp\{ikx + it\sin(\lambda k) + t\cos(\lambda k) + \ln \tilde{y}_0(k)\} \, dk \tag{8}$$

and use the method of steepest descent. The extrema of the real part of the exponent obey

$$-\lambda t \sin(\lambda k) + \frac{d\tilde{y}_0(k)}{dk} \bigg/ \tilde{y}_0(k) = 0 \tag{9}$$

For the assumed form of $\tilde{y}_0(k)$, Eq. (9) is solved by $k=0$. For sufficiently small $t$, $k=0$ is the only solution. As $t$ grows, there may be other solutions [possibly, a sequence of solutions $k_n(\lambda;t)$, the number of which grows with $t$]. Hence, the following analysis is confined to sufficiently short times, so that only $k=0$ is a solution. The second derivative of the exponent at $k=0$ is

$$-\lambda^2 t - \alpha$$

$$\left(\alpha = -\frac{d^2 \tilde{y}_0(k)}{dk^2} \bigg/ \tilde{y}_0(k) \bigg|_{k=0} > 0\right) \tag{10}$$

In this approximation, using $\sin(\lambda k) \approx \lambda k$ in Eq. (9), the latter yields for small $t$ and $x$:

$$y(t,x) \cong \tilde{y}_0(k=0) \frac{1}{\sqrt{\alpha + \lambda^2 t}} \exp\left[-\frac{1}{2} \frac{(x + \lambda t)^2}{(\alpha + \lambda^2 t)}\right] \tag{11}$$

Using Eq. (1), the approximate expression for $u(t,x)$ becomes

$$u(t,x) \cong -\frac{(x + \lambda t)}{(\alpha + \lambda^2 t)} \xrightarrow{t \ll (\alpha/\lambda^2)} -\frac{1}{\alpha}(x + \lambda t) \tag{12}$$

For $t \ll (\alpha/\lambda^2)$, this approximate solution describes a wave that propagates with a velocity $-\lambda$. It can be shown that, under the same assumptions and for small $t$, Eq. (5) also yields the result of Eq. (12).

In Fig. 1, the numerical results for $u(t,x)$ computed through Eqs (8) and (1) are presented for a range of $t$, and a bounded range in $x$, where the approximation of Eq. (12) is expected to hold. The initial data have a Gaussian spectrum in $k$-space:

$$\tilde{y}_0(k) = \exp\left[-\left(k^2/2\sigma^2\right)\right] \tag{13}$$

Clearly, depending on the values of $\alpha$ and $\lambda$, the simple wave structure will not hold for all $t$ and $x$, as dispersion will affect the form of the solution. We demonstrate this in Fig. 2, for the same profile [Eq. (13)] and the same parameter values as in Fig.1.

Eq. (4) also has exact moving-front wave solutions characteristic of the Burgers equation [17]. We look for solutions of the form

$$y(t,x) = y(\xi) \qquad \xi = x + Vt \tag{14}$$

Substituting Eq. (14) in Eq. (2), the latter becomes

$$V y_\xi(\xi) = y(\xi + \lambda) - y(\xi) \tag{15}$$

For a solution of the form

$$y(\xi) = \exp(\rho \xi) \tag{16}$$

the parameter $\rho$ has to obey the dispersion relation

$$V \rho = \exp(\rho \lambda) - 1 \tag{17}$$

Eq. (17) has infinitely many complex solutions. We concentrate on the real ones, as they generate moving front solutions that do not decay. For $(V/\lambda)<0$, and for $(V/\lambda)=1$, Eq. (17) is solved only by $\rho=0$, so that no front solution is possible. The interesting case is that of $(V/\lambda)>0$, $(V/\lambda)\neq 1$, when Eq. (17) has two real solutions: $\rho = 0$ and $\rho = \rho^* \neq 0$. Then, the general wave-solution of Eq. (14) is

$$y(t,x) = a + b\exp\left(\rho^*(x + Vt)\right) \tag{18}$$

Using Eq. (1), yields for $u(t,x)$ a wave front, which propagates with a velocity, $-V$, and varies between two asymptotic values. For $\rho^*>0$ [corresponding to $(V/\lambda)>1$], it has the behavior

$$u(t,x) = \frac{b\rho^* \exp(\rho^* \xi)}{a + b\exp(\rho^* \xi)} \qquad \begin{cases} \xrightarrow{\xi \to -\infty} & 0 \\ \xrightarrow{\xi \to +\infty} & \rho^* \end{cases} \tag{19}$$

In the case $\rho^*<0$ [corresponding to $(V/\lambda)<1$], the roles of $\xi=+\infty$ and $\xi=-\infty$ are interchanged. Fig. 3 shows typical wave-front solutions.

Eq. (4) is a generator of the whole hierarchy [18-20] of the Burgers equation [17]. For initial data that are $C^\infty$, the Taylor expansion of the r.h.s. of Eq. (4) yields

$$u_t = \sum_{n=1}^{\infty} \frac{\lambda^n}{n!} S_n[u] \tag{20}$$

The first three terms in the sum are given below

$$\begin{aligned} S_1[u] &= u_x \\ S_2[u] &= 2u\,u_x + u_{xx} \\ S_3[u] &= 3u^2\,u_x + 3u\,u_{xx} + 3u_x^2 + u_{xxx} \end{aligned} \tag{21}$$

Of these, $S_2[u]$ is just the r.h.s. of the original Burgers equation, and $S_3[u]$ is the next member of the hierarchy.

We comment in passing that Eq. (21) may be viewed as a normal form [23-26] for a Burgers equation with a particular structure of higher-order perturbations.

That the expansion, indeed, yields the whole Burgers hierarchy can be seen through the Taylor expansion of the r.h.s. of the second member of the Lax pair, Eq. (2):

$$y_t = \sum_{n=1}^{\infty} \frac{\lambda^n}{n!} \partial_x^n y(t,x) \tag{22}$$

Each term in the sum on the r.h.s. of Eq. (22) is a member of the well-known sequence of Lax pairs that generates the Burgers hierarchy, comprised of Eq. (1) and of

$$y_t = \partial_x^n y(t,x) \tag{23}$$

The condition for the consistency of Eqs. (1) and (23) is the $n$'th order member of the Burgers hierarchy:

$$u_t = S_n[u] \tag{24}$$

These appear in analyses of the perturbed Burgers equation in the context of the integrability of PDE's [21–23], or perturbative expansions using the Method of Normal Forms [23-27] or the Method of Multiple Time Scales [28–34].

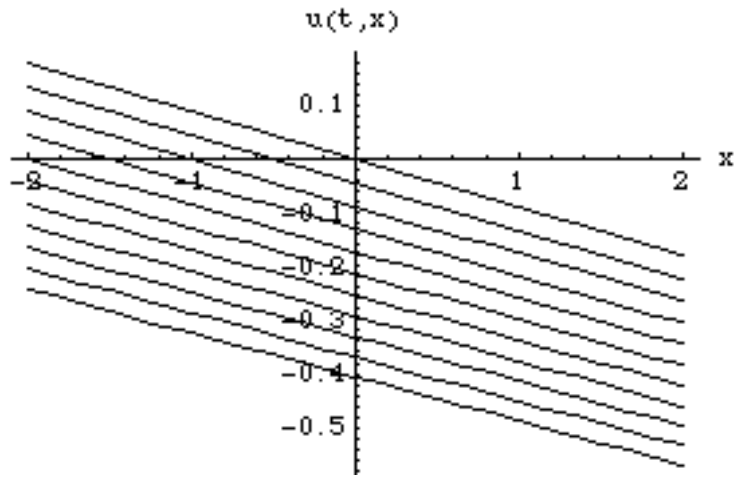

Fig. 1: $u(t,x)$, solved through Eqs. (1), (8) & (13), with $\sigma=0.3$, $\lambda=0.25$, plotted against $x$, for $t$ varying (downwards) from 0 to 20 in steps of 2.

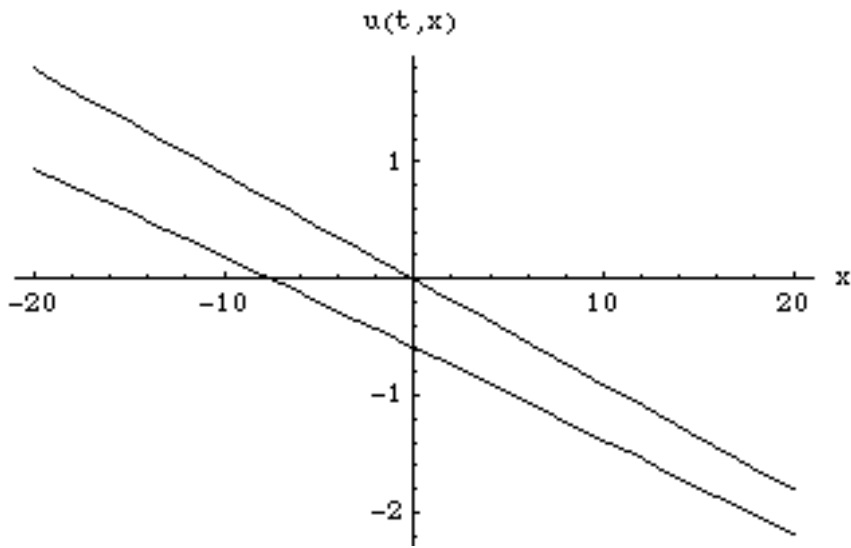

Fig. 2: $u(t,x)$, solved through Eqs. (1), (8) & (13), with $\sigma=0.3$, $\lambda=0.25$, plotted against $x$, for $t=0$ (upper curve) and $t=30$ (lower curve).

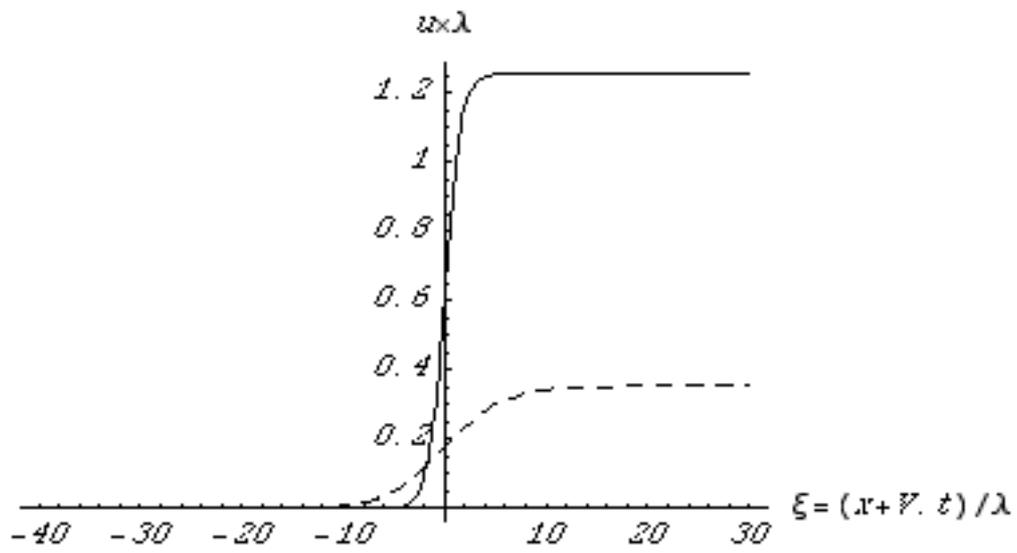

Fig. 3: Wave-front solutions of Eq. (4) of the form of Eq. (19) with λ = 0.25 and V = 0.5 (full line), V = 0.3 (dashed line).